\documentclass[pdflatex,sn-mathphys-num]{sn-jnl}

\usepackage{graphicx}%
\usepackage{multirow}%
\usepackage{amsmath,amssymb,amsfonts}%
\usepackage{amsthm}%
\usepackage{mathrsfs}%
\usepackage[title]{appendix}%
\usepackage{xcolor}%
\usepackage{textcomp}%
\usepackage{manyfoot}%
\usepackage{booktabs}%
\usepackage{algorithm}%
\usepackage{algorithmicx}%
\usepackage{algpseudocode}%
\usepackage{listings}%
\usepackage{amssymb}
\usepackage{hyperref}
\usepackage{bm}

\theoremstyle{thmstyleone}%
\theoremstyle{thmstyletwo}%

\theoremstyle{thmstylethree}%

\begin{document}

\title[Article Title]{EAC-Net: Predicting real-space charge density via equivariant atomic contributions}


\author[1,2,4]{\fnm{Xuejian} \sur{Qin}}\email{qinxuejian@nimte.ac.cn}
\equalcont{These authors contributed equally to this work.}

\author[2,3]{\fnm{Taoyuze} \sur{Lv}}\email{taoyuze.lv@ustc.edu.cn}
\equalcont{These authors contributed equally to this work.}

\author*[2,3,5]{\fnm{Zhicheng} \sur{Zhong}}\email{zczhong@ustc.edu.cn}

\affil[1]{\orgdiv{Key Laboratory of Marine Materials and Related Technologies, Zhejiang Key Laboratory of Marine Materials and Protective Technologies}, \orgname{Ningbo Institute of Materials Technology and Engineering, Chinese Academy of Sciences}, \orgaddress{\city{Ningbo}, \country{China}}}

\affil[2]{\orgdiv{Suzhou Institute for Advanced Research}, \orgname{University of Science and Technology of China}, \orgaddress{\city{Suzhou}, \country{China}}}

\affil[3]{\orgdiv{School of Artificial Intelligence and Data Science}, \orgname{University of Science and Technology of China}, \orgaddress{\city{Hefei}, \country{China}}}

\affil[4]{\orgdiv{College of Materials Science and Opto-Electronic Technology}, \orgname{University of Chinese Academy of Sciences}, \orgaddress{\city{Beijing}, \country{China}}}

\affil[5]{\orgdiv{Suzhou Lab}, \orgaddress{\city{Suzhou, Jiangsu}, \country{China}}}

\abstract{Charge density is central to density functional theory (DFT), as it fully defines the ground-state properties of a material system. Obtaining it with high accuracy is a computational bottleneck. Existing machine learning models are constrained by trade-offs among accuracy, efficiency, and generalization. Here, we introduce the Equivariant Atomic Contribution Network (EAC-Net), which couples atoms and grids to integrate the strengths of grid-based and basis-function frameworks. EAC-Net achieves high accuracy (typically below 1\% error), enhanced efficiency, and strong generalization across complex systems. Building on this framework, we develop EAC-mp, a universal charge density model covering the periodic table. The model demonstrates robust zero-shot performance across diverse systems, and generalizes beyond the training distribution, supporting downstream applications such as band structure calculations. By linking local chemical environments to charge densities, EAC-Net provides a scalable framework for accelerating electronic structure prediction and enabling high-throughput materials discovery.}




\maketitle

\section{Introduction}\label{sec1}

Charge density lies at the heart of density functional theory (DFT), uniquely determining all ground-state properties of a material system\cite{PhysRev.140.A1133,PhysRev.136.B864}. However, accurate evaluation of charge density requires solving the Kohn-Sham equations self-consistently, which can be prohibitively expensive for large-scale or high-throughput simulations. This motivates the development of efficient methods to predict charge density without explicitly solving the full quantum problem. In recent years, deep learning methods have emerged as a promising alternative to conventional DFT calculations, offering the potential to predict energies, interatomic forces\cite{batatia2024foundationmodelatomisticmaterials,yang2024mattersimdeeplearningatomistic,zengDeePMDkitV3MultipleBackend2025,fu2025learningsmoothexpressiveinteratomic,kimDataEfficientMultifidelityTraining2025} as well as charge densities\cite{kokerHigherorderEquivariantNeural2024a,PhysRevB.108.235159,jorgensenEquivariantGraphNeural2022,ZEPEDANUNEZ2021110523,PhysRevB.110.184106,chandrasekaranSolvingElectronicStructure2019a,focassioLinearJacobiLegendreExpansion2023a,lewisLearningElectronDensities2021,grisafiTransferableMachineLearningModel2019,NEURIPS2024_12c118ef,C9SC02696G} with high accuracy and efficiency. Existing deep learning approaches for charge density prediction can be broadly categorized into grid-based\cite{kokerHigherorderEquivariantNeural2024a,PhysRevB.108.235159,jorgensenEquivariantGraphNeural2022,ZEPEDANUNEZ2021110523,PhysRevB.110.184106,chandrasekaranSolvingElectronicStructure2019a,focassioLinearJacobiLegendreExpansion2023a} and basis-function\cite{lewisLearningElectronDensities2021,grisafiTransferableMachineLearningModel2019,NEURIPS2024_12c118ef,C9SC02696G} methods. Grid-based methods are generally more versatile and require minimal prior domain knowledge, but they tend to suffer from reduced computational efficiency. In contrast, basis-function methods have demonstrated higher efficiency\cite{NEURIPS2024_12c118ef}, albeit requiring manual basis function design. While grid-based and basis-function models each offer distinct advantages, it remains an open challenge to combine their strengths. To this end, we seek a formulation that is both physically grounded and computationally efficient.


Inspired by these models, we propose the Equivariant Atomic Contribution Network (EAC-Net) to combine the advantages of the two approaches, predicting the full real-space charge density while implicitly modeling atomic contributions.
Our model leverages the concept of atomic influence as a structural prior to enhance the fidelity and generalizability of charge density predictions. EAC-Net has been tested in various material systems, including crystals, surfaces, amorphous structures, alloys, and molecules. The results demonstrate that EAC-Net can accurately predict charge density with high efficiency, and also provide a learned charge partition scheme. The EAC framework can not only assist the DFT electronic structure calculations but also be used to analyze charge-related properties of materials, such as charge transfer, catalytic activity, and reactivity. 

Leveraging the dataset from Materials Project (MP)\cite{Munro2020, Merchant2023}, we further develop EAC-mp as a large-scale variant that delivers high accuracy while being computationally efficient and providing atomic-decomposed charge densities. The model demonstrates exceptional zero-shot prediction capabilities across diverse material systems, offering a new solution for high-throughput materials screening.

\section{Results}\label{sec2}

\subsection{Model architecture}\label{subsec2}

\begin{table}[h]
\caption{Comparison among charge density prediction models}\label{tab_1}
\begin{tabular*}{\textwidth}{@{\extracolsep\fill}lccc}
\toprule
Methods & Grid-based & Basis-function & Atom-grid coupling \\
\midrule
Core idea & Decode from grid & Basis superposition & Decode from atom-grid edge \\
$\rho=$ &  $\mathrm{Net}(D_{\mathrm{grid}})$  & $\sum_i^\mathrm{atom}\sum_k^\mathrm{basis}c_{ik}f_k$ & $\sum_i^\mathrm{atom}\mathrm{Net}(D_i)$\\
Atom contrib.    & $\times$ & \checkmark & \checkmark  \\
Application   & All & Molecules\footnotemark[1] & All\\
Efficiency    & Low  & High & High \\
Refs.       & ChargE3Net, etc.\cite{kokerHigherorderEquivariantNeural2024a,PhysRevB.108.235159,jorgensenEquivariantGraphNeural2022,ZEPEDANUNEZ2021110523,PhysRevB.110.184106,chandrasekaranSolvingElectronicStructure2019a,focassioLinearJacobiLegendreExpansion2023a} & SCDP, etc.\cite{lewisLearningElectronDensities2021,grisafiTransferableMachineLearningModel2019,NEURIPS2024_12c118ef,C9SC02696G} & This work         \\
\botrule
\end{tabular*}

\footnotetext[1]{Ref. \cite{lewisLearningElectronDensities2021} have implemented basis-function method in condensed phases, such as H$_2$O, Al, and Si.}
\end{table}

As listed in Table \ref{tab_1}, existing deep-learning approaches to charge density prediction differ primarily in how they represent the underlying electronic distribution. Grid-based methods\cite{kokerHigherorderEquivariantNeural2024a,PhysRevB.108.235159,jorgensenEquivariantGraphNeural2022,ZEPEDANUNEZ2021110523,PhysRevB.110.184106,chandrasekaranSolvingElectronicStructure2019a,focassioLinearJacobiLegendreExpansion2023a} are ``grid-focused''. They first construct descriptors at each grid point from the surrounding atomic environment, and then decode these descriptors to obtain the local charge density. However, the descriptors at grid points are inherently nonlinear due to the overlapping contributions from neighboring atoms, which prevents their direct use for predicting partial charge densities. In contrast, basis-function methods\cite{lewisLearningElectronDensities2021,grisafiTransferableMachineLearningModel2019,NEURIPS2024_12c118ef,C9SC02696G} are ``atom-focused''. They learn the expansion coefficients of the charge density in a predefined atomic basis, from which the continuous density is reconstructed in space. These methods are usually used in organic molecules. Achieving higher accuracy often requires large basis sets, which may include virtual orbitals in addition to atomic orbitals\cite{NEURIPS2024_12c118ef}. Although physically grounded, this representation requires carefully designed and often system-specific basis sets, which limit scalability to diverse materials systems.


\begin{figure}[h]
\centering
\includegraphics[width=1.0\textwidth]{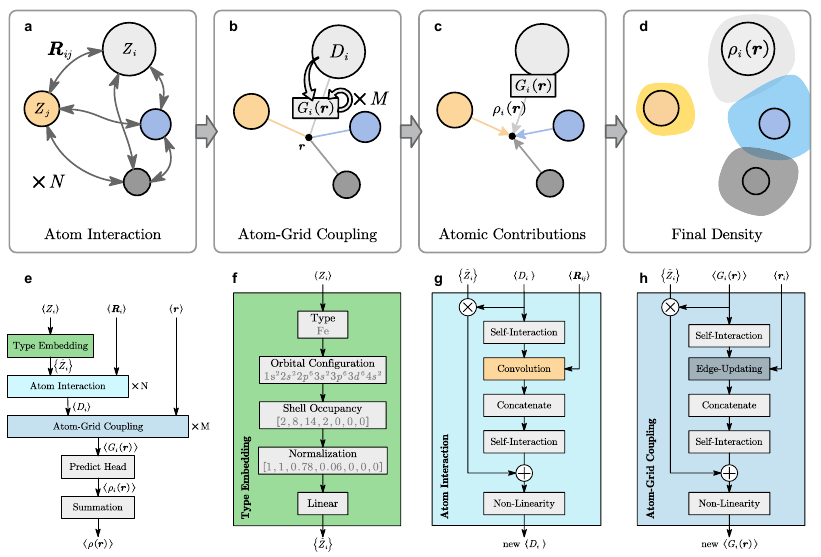}
\caption{Workflow and the architecture of the EAC-Net. \textbf{a}-\textbf{d}, Schematics of message passing process for the atom-grid coupling strategy. \textbf{e}, The overall architecture of EAC-Net. \textbf{f}-\textbf{h}, The architectures of type embedding, atom interaction, and atom-grid coupling blocks, respectively. To improve visual clarity, less critical subscripts and superscripts ($l$ and $c$) are omitted in the figure.}\label{fig_1}
\end{figure}

Here, the EAC strategy aims to combine the advantages of both grid-based and basis-function approaches while mitigating their respective limitations. This hybrid design removes the need for manual basis set design, while providing a general framework capable of accurately predicting charge density across diverse systems. The overall workflow of EAC-Net is illustrated in Fig. \ref{fig_1}a-d and consists of four main stages. First, the atom interaction block embeds local environment information into equivariant descriptors for each atom. Second, the atom-grid coupling block maps atomic descriptors onto atom-grid edges, enabling the construction of edge features that capture directional and local contributions to the charge density. Third, the decoding step transforms edge features into atomic contributions. Finally, the aggregation stage sums contributions from all atoms to obtain the total charge density. This hybrid design enables atomic descriptors to be computed once and reused for arbitrary grid resolutions, while preserving directional and local detail.

Fig. \ref{fig_1}e depicts the overall architecture of EAC-Net. In the following, we provide a detailed description of each component.

\textbf{Type embedding.} For each atom $i$, the atomic type $Z_i$ is first embedded into $\tilde{Z}_i$ through a type encoding block. EAC-Net supports multiple encoding schemes, including one-hot encoding and shell-filling encoding. As shown in Fig. \ref{fig_1}f, the shell-filling approach first extracts the orbital configurations from atom types. Then a shell occupancy array is constructed for each atom. These arrays are then normalized by the maximum occupancy of each orbital, and then projected into the embedding space via a linear transformation. These encoding strategies can be combined by concatenation to form a richer representation, enabling the model to incorporate both discrete and electronic structural information in $\tilde{Z}_i$. 

\textbf{Atom interaction.} The type embedding vectors are input into the atom interaction block and obtain equivariant descriptors (Fig. \ref{fig_1}g), using an E(3)-equivariant architecture implemented with the \texttt{e3nn} framework \cite{geiger2022e3nneuclideanneuralnetworks}, adopting the NequIP architecture \cite{batznerE3equivariantGraphNeural2022} to construct descriptors $D^l_{ic}$, where $c$ denotes the channel index and $l$ denotes the angular momentum order.

\textbf{Atom-grid coupling.} The major difference between EAC and other models is the atom-grid coupling block. Once the atom-interaction block is complete, each atom is equipped with an equivariant representation, enabling the subsequent decoding of the charge density. As shown in Fig. \ref{fig_1}h, this decoding process is carried out by the atom-grid coupling block, which shares a similar structure with the atom interaction block but replaces the convolution layer with an edge-updating layer. Specifically, for each atom-grid pair, the edge feature is updated as:
\begin{equation}
    \mathcal{E}^{l_o}_{ic}\left(\bm r,G_{ic}^{l_q}(\bm r)\right) = F_{c}^{l_pl_q}(\bm{r})\otimes_{l_p,l_q \to l_o}G_{ic}^{l_q}(\bm r)
\end{equation}
where $\otimes_{l_p,l_q \to l_o}$ is the tensor product (see \ref{method:model} in Method for details). In this formulation, information is propagated from atomic nodes to edges that connect with spatial grids through a tensor product with the embedded relative position, preserving equivariance. Unlike conventional grid-based methods that aggregate atomic information to construct grid-point descriptors, our approach does not rely on any grid-point aggregation. Instead, it directly models the interaction between each atom and grid point via edge features $G_{ic}^{l}$, preserving directional detail and avoiding redundancy.

\textbf{Atomic contribution.} The third step involves decoding the partial charge density $\rho_i(\bm{r})$ from these edge features. The scalar features ($l=0$) of input $G_{ic}^{l}$ are first compressed through a bottleneck layer. To enrich the information content of the resulting representation, additional features such as $r_{i}$ (the relative position between atom $i$ and grid point at $\bm r$), $B(r_{i})$, and $Z_i$ can be optionally concatenated to it, depending on the specific task requirements: 
\begin{equation}
    G_{i}(\bm r) = \mathrm{MLP}(G_{ic}^0(\bm r)) | \big|_{f\in \mathcal{S}}f, \mathcal{S}\subseteq \left\{r_{i}, B(r_{i}), Z_i\right\}
    \label{eq_cat}
\end{equation}
where $\mathcal{S}$ is any subset of the additional features, including the empty set, and all of the selected features are concatenated together. Then the $\rho_i(\bm{r})$ is decoded from $G_{i}(\bm r)$ via:
\begin{equation}
    \rho_i(\bm{r}) = v\left(G_{i}(\bm r)\right) \cdot w\left(G_{i}(\bm r), r_{i}\right)
    \label{eq_head}
\end{equation}
Here, the scalar edge features are passed through two separate networks: value net $v$ and weight net $w$. The output of $v$ can be interpreted as the candidate values contributed by each atom, while $w$ is flexible and can take various forms, depending on how edge features, geometric distances, and neighboring edge information are integrated. We explore several instantiations of $w$, including combinations of smooth functions and softmax techniques. Detailed formulations and explanations for each variant are provided in Methods. This formulation introduces a simple attention mechanism in which the contribution of each atom to the charge density at a spatial point $\bm{r}$ is weighted based on its relative importance. 

\textbf{Summation.} The predicted total charge density $\hat{\rho}(\bm{r})$ is then obtained by adding the atomic contributions from all atoms:
\begin{equation}
    \hat\rho(\bm{r})=\sum_{i=1}^{N_a}\rho_i(\bm{r})
    \label{eq_1}
\end{equation}

Although the atom-interaction stage of EAC-Net shares a similar backbone with prior equivariant architectures, the subsequent coupling-decoding procedure fundamentally changes how real-space properties are represented. In conventional grid-based approaches, descriptor construction at each spatial location leads to considerable computational and memory overhead. In EAC-Net, the atomic descriptors from the interaction stage can be computed once and cached, and reused for subsequent density evaluations at arbitrary spatial resolutions or sampling patterns. Only the atom-grid coupling step needs to be recomputed, as it depends on the relative positions between atoms and spatial points. This design reduces redundant computation, improves memory efficiency, and makes the model more practical for repeated or high-resolution density evaluations.

\subsection{Model performance}

\begin{figure}[h]
\centering
\includegraphics[width=1.0\textwidth]{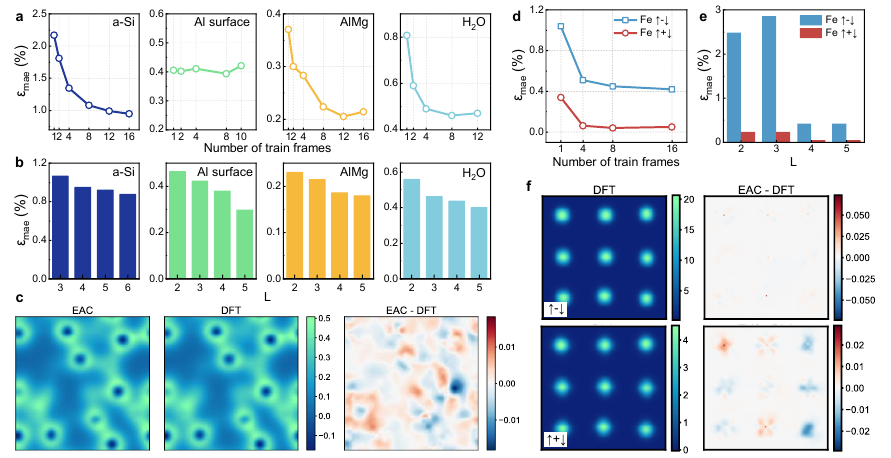}
\caption{Accuracy evaluation of EAC-Net across amorphous silicon, aluminum surface, Al-Mg alloy, water, and spin-polarized Fe. \textbf{a}, Convergence of model accuracy as a function of the number of structural frames used for training. \textbf{b}, Dependence of model accuracy on the angular momentum cutoff $L$. \textbf{c}, A comparison between the charge density of the amorphous Si from EAC-Net and DFT is shown, with the 2D slice exhibiting the area of largest error in the prediction. \textbf{d} and \textbf{e}, The convergence tests for spin-polarized density of fcc Fe. \textbf{f}, The real and error distributions for total and spin densities. }\label{fig_2}
\end{figure}

EAC-Net was trained and validated on a range of complex atomic systems, including amorphous silicon, aluminum surfaces, Al-Mg alloys, and liquid water. To evaluate predictive accuracy, we tested the model's performance with respect to both the number of training frames and the angular momentum cutoff $L$. The normalized mean absolute error $\varepsilon_{\mathrm{mae}}$ was used as the evaluation metric\cite{kokerHigherorderEquivariantNeural2024a,NEURIPS2024_12c118ef}:
\begin{equation}
    \varepsilon_{\mathrm{mae}}=\frac{\int|\rho(\bm r)-\hat\rho(\bm r)|d\bm r}{\int|\rho(\bm r)|d\bm r}
\end{equation}
where the integration is performed over all spatial grid points.

As shown in Fig. \ref{fig_2}a, EAC-Net achieves convergence with a remarkably small number of training frames for all systems tested, typically between 8 and 12. Notably, for the aluminum surface, accurate prediction is achieved with just one DFT frame. In the $L$-dependence studies (Fig. \ref{fig_2}b), the $\varepsilon_{\mathrm{mae}}$ decreases consistently with increasing $L$, indicating that higher-order angular features significantly improve model performance, particularly in complex systems with covalent or metallic bonding. This trend is consistent with findings from prior work\cite{kokerHigherorderEquivariantNeural2024a}. However, a larger $L$ significantly increases the computational burden, leading to longer training and inference times. Therefore, a trade-off between accuracy and computational efficiency should be carefully considered before training. 

The model demonstrates excellent overall accuracy across complex systems, consistently achieving $\varepsilon_{\mathrm{mae}}$ values below 1\%. Among the four tested systems, the Al-Mg alloy yields the highest accuracy, with $\varepsilon_{\mathrm{mae}}$ dropping below 0.2\%. In contrast, the performance on amorphous silicon is comparatively lower, with errors approaching 0.8\%. The errors generally occur around Si atoms and bond positions (Fig. \ref{fig_2}c). This decrease in accuracy is primarily attributed to the structural disorder inherent in amorphous systems, which introduces highly complex local atomic environments. Moreover, unlike water, silicon tends to form covalent-bonding networks, leading to a substantial deviation of the charge density from spherical symmetry in real space. However, these challenges can be mitigated by enlarging the training dataset and extending the training duration. Overall, the results indicate that for charge density prediction tasks, the training difficulty is more sensitive to structural disorder than to compositional disorder, and ordered systems are the easiest for the model to learn.

Beyond the total charge density (spin-up plus spin-down), electrons also possess the spin degree of freedom. We further evaluated the model's capability to predict the spin density, defined as the difference between spin-up and spin-down electron densities. Although this may appear as simply introducing an additional channel in the output, the underlying spin density often exhibits sharper spatial features and higher-order angular structures, particularly in systems containing transition metals or magnetic ordering. These characteristics make the prediction task significantly more challenging for neural networks.

As shown in Fig. \ref{fig_2}d and e, EAC-Net achieves high accuracy in predicting the total charge density, with an $\varepsilon_{\mathrm{mae}}$ of just 0.048\% for the $L=4$ model. The prediction errors are localized near atomic centers, as highlighted in Fig. \ref{fig_2}f. In contrast, the spin density prediction exhibits an order of magnitude higher error, with $\varepsilon_{\mathrm{mae}}$ around 0.42\%,  reflecting the difficulty of capturing its finer spatial features. While the total charge density around each Fe atom remains nearly spherical, the corresponding spin density exhibits a distinct octahedral symmetry. Despite the reduced accuracy, the model is still able to qualitatively reproduce the high-frequency, symmetry-breaking features of the spin distribution, highlighting its potential applicability to spin-polarized and magnetic systems.

\begin{figure}[h]
\centering
\includegraphics[width=0.5\textwidth]{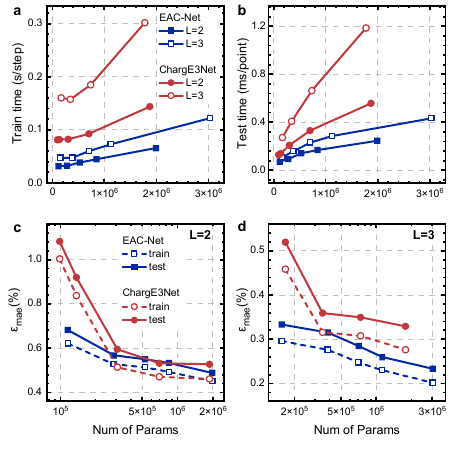}
\caption{The performance comparisons between EAC-Net and ChargE3Net in crystalline silicon. \textbf{a}, Plot for the time per step of training with respect to different model sizes and $L$. \textbf{b}, The inference time per grid point on a whole structure. \textbf{c} and \textbf{d}, The model scaling plots for $L=2$ and $L=3$, respectively.}\label{fig_compare}
\end{figure}

Fig. \ref{fig_compare} presents a comparative analysis of the performance of EAC-Net and ChargE3Net, trained on the same crystalline silicon dataset across various model sizes for 150,000 training steps. As shown in Fig. \ref{fig_compare}a, EAC-Net requires two to four times less training time than ChargE3Net for a comparable number of parameters. Moreover, the growth rate of inference time with respect to model size is significantly lower for EAC-Net than for ChargE3Net (Fig. \ref{fig_compare}b). Notably, EAC-Net also mitigates the efficiency degradation associated with increasing angular resolution $L$. For instance, when $L=3$, the training and inference times of EAC-Net increase by less than 1.5 times compared to the $L=2$ case, whereas for ChargE3Net, the corresponding increase is approximately twofold. 

The enhanced efficiency of EAC-Net in inference can be attributed to its atom descriptor caching mechanism. During inference, the atom interaction block is computed only once for all atoms. This precomputed information is then reused for predictions on spatial grids, enabling efficient atom-grid coupling without redundant computation. It is worth noting, however, that ChargE3Net could, in principle, achieve similar efficiency improvements through the adoption of a tailored caching strategy. The potential benefits of such a mechanism for ChargE3Net and other models merit further investigation.

As model size increases, both EAC-Net and ChargE3Net exhibit improved predictive accuracy (Fig. \ref{fig_compare}c and d). Importantly, the accuracy gain with increasing model size is more pronounced in the $L=3$ setting compared to $L=2$, suggesting that higher-order angular features enhance the model's representational capacity and contribute to more favorable scaling behavior. For models with more than 200,000 parameters in the $L=2$ setting, EAC-Net performs comparably to ChargE3Net. In contrast, for $L=3$, EAC-Net consistently outperforms ChargE3Net across all tested parameter scales. Furthermore, the scaling trend of accuracy with respect to model size is more favorable for EAC-Net.

\begin{figure}[h]
\centering
\includegraphics[width=0.88\textwidth]{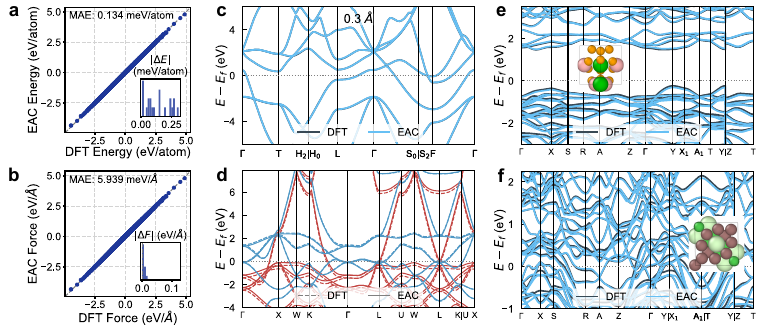}
\caption{Non-self-consistent field DFT calculations from EAC charge densities. \textbf{a} and \textbf{b}, The comparison of energies and forces of Si predicted from EAC and DFT charge densities. \textbf{c}, The band structure of Si under 0.3 \AA~atomic displacement, calculated from DFT (grey line) and EAC-Net (light blue line) charge densities. \textbf{d}, The comparison of spin-polarized band structures of Fe. The blue lines indicate the spin-down state and the red lines are the spin-up state.  \textbf{e} and \textbf{f}, Band structures of two randomly chosen MP structures from the test dataset of EAC-mp ($\mathrm{Pr_5In_{11}Ni_6}$ and $\mathrm{Ba(BSe_3)_2}$). }\label{fig_band}
\end{figure}

To evaluate both the predictive accuracy and extrapolation capability of the model, we performed non-self-consistent DFT calculations using charge densities predicted by the model. We used Si (supercell with 64 atoms) with perturbations of 0.05 and 0.1 \AA~to examine the accuracy of energy and force calculations. The model had only been trained on structures with atomic positions perturbed up to 0.1 \AA. As shown in Fig. \ref{fig_band}a and b, the EAC-based calculations achieved mean absolute errors (MAEs) of 0.134 meV/atom for total energy and 5.939 meV/\AA~for atomic forces, respectively. These results not only demonstrate high fidelity in reproducing DFT-level predictions but also significantly outperform typical machine learning force fields developed for silicon systems\cite{li_unified_2020,https://doi.org/10.1002/cphc.202400090}, which typically do not rely on explicitly learned electronic structure information. These findings validate the model's ability to produce physically accurate predictions across a realistic configuration space.

We further calculate the band structures for comparison. As an initial test case, we considered cubic silicon (space group $\mathrm{Fd\bar{3}m}$). To introduce a nontrivial perturbation, displacements of 0.3 \AA~were applied to the central silicon atom. As shown in Fig. \ref{fig_band}c, the resulting band structures derived from the predicted charge densities exhibit excellent agreement with the DFT ground-truth results. Any visible discrepancies are limited to a few localized regions, barely distinguishable from the reference. Band comparison results under other displacements and tests in cubic GaAs ($\mathrm{F\bar{4}3m}$) are also listed in Supplementary Information. To probe transferability under spin effects, we further examined spin-polarized bcc-Fe. The predicted spin-up and spin-down bands of bcc-Fe (Fig. \ref{fig_band}d) show excellent agreement with DFT, confirming the model's applicability to magnetic systems. Finally, we include tests using EAC-mp, our universal charge-density predictor, on two previously unseen structures. Detailed assessments of EAC-mp are presented in the following section. The predicted band structures remain in close agreement with DFT, with deviations slightly larger than those of the task-specific model but still confined to small, localized regions. Collectively, these results establish that EAC-Net not only delivers high accuracy on perturbed and magnetic systems but also provides a solid foundation for universal charge-density prediction via EAC-mp.

\subsection{Generalization potential and emergent behaviors}

\begin{figure}[h]
\centering
\includegraphics[width=1\textwidth]{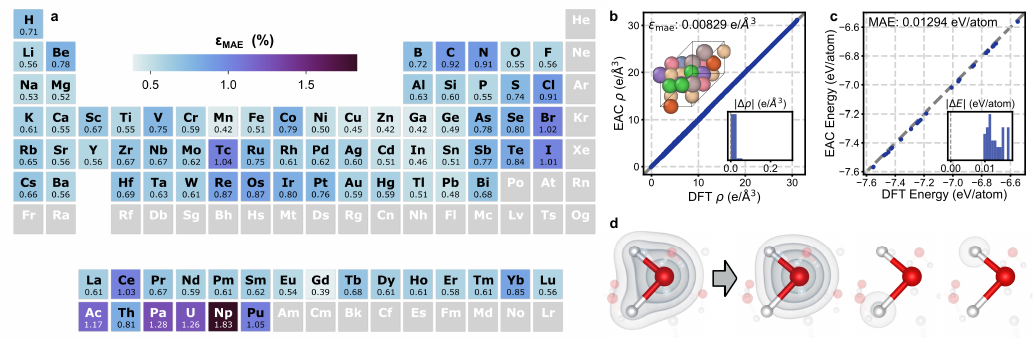}
\caption{The accuracy and generalization ability of the EAC-mp model. \textbf{a}, The average $\varepsilon_{\mathrm{mae}}$ across the periodic table. \textbf{b}, The comparison between DFT total charge density and EAC-mp zero-shot prediction for high-entropy alloys containing Al, Co, Cr, Fe, Mn, Ni, and Si. \textbf{c}, The comparison between energies of perturbed high-entropy alloy structures calculated from DFT and predicted charge density. \textbf{d}, The atom-wise decomposition of charge density in H$_2$O.}\label{fig_5}
\end{figure}

Having established the accuracy and reliability of the EAC-Net architecture through targeted evaluations on various complex systems, we next explored its potential as a general-purpose charge density predictor. To this end, we trained a larger model, termed EAC-mp, on a chemically diverse subset of the Materials Project database. Despite the limited size of the training set, the goal was to test the model’s ability to learn transferable representations of charge density that generalize across element types and structural motifs. We then evaluated EAC-mp on over 20,000 frames drawn from materials outside the training distribution. For each frame, we computed the $\varepsilon_{\mathrm{mae}}$ of charge densities, and summarized the average $\varepsilon_{\mathrm{mae}}$ for each element in Fig. \ref{fig_5}a.

The model achieves $\varepsilon_{\mathrm{mae}}$ values below 1.0\% for the vast majority of chemical elements. Notably, only a small number of elements—primarily those that are radioactive, sparsely represented in the dataset, or chemically exotic—exhibit $\varepsilon_{\mathrm{mae}}$ exceeding 1.0\%. These include actinides such as Np, U, and Pa, for which limited training data likely contribute to increased prediction errors.

Importantly, the distribution of errors across the periodic table appears remarkably uniform; most of the elements yield errors around 0.6\%, with no pronounced clusters of high or low accuracy corresponding to specific chemical groups or periods. For instance, both main-group elements and transition metals achieve comparable performance, and no particular family dominates the tail of the error distribution. This suggests that the model is not overfitting to particular bonding types or chemical environments, and instead has learned robust, transferable representations of charge density applicable across a wide range of stoichiometries.

This uniformity in predictive performance, combined with the relatively low global error, provides strong empirical evidence for the generalization capability of EAC-mp. The model is not only capable of interpolating within familiar regions of chemical space, but also exhibits stable behavior when extrapolated to less common elements.

In addition to the standard test set, we further examined the zero-shot generalization performance of EAC-mp on chemically and structurally complex high-entropy alloys (HEAs) containing Al, Co, Cr, Fe, Mn, Ni, and Si. These systems, characterized by significant compositional and configurational disorder, were not included in the training set and thus represent a stringent test of model transferability. For each HEA configuration, we performed two separate evaluations. First, we compared the predicted charge densities with reference values obtained from unperturbed (relaxed) geometries, assessing the model's ability to reproduce physically meaningful electron distributions without prior exposure to similar systems (\ref{fig_5}b). Second, we used the predicted charge densities of EAC-mp to calculate non-self-consistent total energies on HEA structures subjected to small atomic displacements (0.05 \AA), thus probing the model's sensitivity to subtle configurational variations (\ref{fig_5}c).

The results from both evaluations indicate that EAC-mp retains high fidelity in electronic structure prediction, even under conditions of pronounced chemical complexity. We also observed a small but consistent energy offset, which is likely due to the MP charge densities being spin-polarized, whereas our calculations omitted spin polarization to avoid the complexity and uncertainty of such treatments. These findings underscore the model's capacity to generalize effectively across domains, offering a scalable approach for electronic property estimation in materials beyond the training distribution.

Moreover, EAC-Net naturally provides atom-wise contributions to the total charge density, as illustrated by the water molecule in Fig. \ref{fig_5}d, where the total density is automatically decomposed into contributions from hydrogen and oxygen atoms. In conventional approaches, such atom-resolved information typically requires post-processing with classical partitioning schemes such as Mulliken\cite{mullikenElectronicPopulationAnalysis1955} and Atoms in Molecules (AIM)\cite{10.1093/oso/9780198551683.001.0001} methods. While the quantitative and qualitative validity of these decompositions remains to be fully established, the atom-wise outputs of EAC-Net offer a direct and flexible representation that could be further refined by incorporating additional physical or experimental constraints. Together, these results suggest that EAC-Net performs well on system-specific predictions and holds promise as a scalable and interpretable framework for universal charge density modeling.

\section{Discussion}\label{sec12}

In this work, we introduce the Equivariant Atomic Contribution Network (EAC-Net), a novel framework for accurate and efficient charge density prediction in atomistic systems. By decomposing the total charge density into equivariant atomic contributions and leveraging atom-grid coupling, EAC-Net bridges the gap between grid-based and basis-function approaches. This unique formulation enables accurate reconstruction of real-space charge densities while naturally providing a partitioning scheme that can capture anisotropic and environment-dependent electronic features.

Compared with existing deep learning models, EAC-Net offers several key advantages. First, it achieves high accuracy with remarkably few training samples, making it suitable for regimes with scarce data such as surface science, defect chemistry, and high-throughput screening. Second, it scales favorably with model size and angular resolution, maintaining computational efficiency through its descriptor caching mechanism. Third, it generalizes across a wide spectrum of structural motifs, including crystalline, amorphous, and interfacial systems, and extrapolates reliably under atomic and lattice perturbations. These properties make EAC-Net a versatile and practical tool for both fundamental and applied studies of materials.


Building on these results, we also introduce a preliminary universal model, EAC-mp, trained on a chemically diverse dataset and tested on systems beyond the training distribution. This prototype demonstrates the feasibility of extending the EAC-Net toward a foundation model for charge density prediction, while also indicating that further refinement is required. Future work will expand element coverage and structural diversity, incorporate spin polarization and other electronic states, and improve accuracy and stability, thereby advancing EAC-mp toward a more mature foundation model.

Moreover, a large-scale EAC model could act as a backbone for downstream multi-property prediction, enabling joint learning of quantities such as electrostatic potential, charge transfer integrals, dielectric response, and other properties. It may also serve as a building block for more complex generative or inverse-design frameworks where electronic information plays a central role. Importantly, the modular design of EAC-Net also lends itself naturally to multi-task learning: by adding additional prediction heads on top of the shared atomic descriptors, the model can be extended to jointly predict additional quantum or classical properties in a unified and data-efficient manner.

\section{Methods}\label{sec11}

\subsection{DFT calculations}

All of the DFT calculations in this study were performed using the Vienna Ab initio Simulation Package (VASP)\cite{PhysRevB.54.11169}. We adopted the Perdew-Burke-Ernzerhof (PBE) exchange-correlation functional\cite{PhysRevLett.77.3865} and projected augmented-wave (PAW)\cite{PhysRevB.59.1758} pseudopotentials. Some of the calculation parameters were set with the help of VASPKIT\cite{VASPKIT}. The convergence criterion of the electronic self-consistent loop was set to $10^{-6}$ eV for all single-point calculations. The structural frames of amorphous Si, aluminum surface, and Al-Mg alloy were adopted from the previous study\cite{PhysRevB.108.235159}. For water, we used the SCAN functional with 680 eV plane-wave energy cutoff and 0.5 \AA$^{-1}$ k-point spacing. The structures of crystalline Si (64 atoms) were computed with a manually set $100\times100\times100$ fast Fourier transformation grid. A $4\times4\times4$ Gamma-centered k-point mesh was adopted. For spin-polarized calculations in the Fe crystal (54 atoms), the energy cutoff was set to 520 eV with a $4\times4\times4$ Monkhorst-Pack k-point mesh. Gaussian smearing with a width of 0.1 eV was applied. Magnetism was considered with an initial local magnetic moment of 5 $\mathrm{\mu}$B assigned to each atom.

For band structure calculations, we used the primitive cells of Si or GaAs with perturbation. The high-symmetry k-point path was generated using VASPKIT based on the crystal symmetry. The self-consistent charge density obtained from a preceding static calculation or model prediction was used as input, and non-self-consistent field calculations were performed along the high-symmetry k-path to obtain the band structure.

\subsection{Dataset construction}

Charge density is a dense three-dimensional field, and its spatial smoothness often leads to significant redundancy among neighboring grid points. To improve training efficiency while preserving essential physical features, we adopt a sampling strategy that selectively extracts representative spatial points from each structure. Specifically, we employ a combination of three sampling schemes:

\begin{enumerate}
    \item \textbf{Random sampling with cutoff}: Grid points are randomly sampled within a finite radius centered on each atom. This ensures coverage of both bonding regions and interstitial space, while excluding distant regions beyond the model's cutoff.

    \item \textbf{Density-weighted sampling}: Points are sampled with probabilities proportional to the absolute value of the charge density. This emphasizes regions with strong electronic features, such as atomic cores and bonding areas.

    \item \textbf{Core-focused sampling}: Points are explicitly sampled within a short cutoff radius around atomic nuclei, to better capture steep density gradients and localized variations near atomic centers.
    \item \textbf{Gradient-weighted sampling}: Points are sampled with probabilities proportional to the norm of the gradient of charge densities. This emphasizes regions with high-frequency features.
\end{enumerate}

Each structure's training dataset contains a proportion of points from the above three strategies, providing a balanced representation across different spatial regions.

For the convergence tests of amorphous Si, aluminum surface, Al-Mg alloy, and water, the training frames are limited, and we chose to use the full charge density files as training sets. Meanwhile, for the universal charge density predictor, we used the random sampling strategy to minimize the size of the dataset. For each charge density file, 5,000 grid points were randomly selected.

\subsection{Model details}
\label{method:model}

The key mechanism by which the \verb+e3nn+ framework preserves equivariance is the tensor product operations defined by the \textit{Clebsch-Gordan coefficients}\cite{thomas2018tensorfieldnetworksrotation}. Specifically, the tensor product combines two irreps $x$ and $y$ into a new set of irreps as follows:
\begin{equation}
    \left(
        x^{\,l_p} \otimes_{l_p,l_q \to l_o} y^{\,l_q}
    \right)^{l_o}_{m_o}
    =
    \sum_{m_p, m_q}
    C^{\,l_p l_q l_o}_{m_p m_q m_o} \;
    x^{\,l_p}_{m_p} \;
    y^{\,l_q}_{m_q},
\end{equation}
where $l_p$ and $l_q$ are the angular momentum orders of the input irreps, and $l_o$ indexes the resulting output order. The indices $m_o$, $m_p$, and $m_q$ are the corresponding magnetic quantum numbers. Here, superscripts refer to representation orders, not mathematical exponents. For a given $l$, multiple irreps (i.e., channels) may be present, allowing the model to represent rich and structured features.

For the atom interaction module, the encoded atomic type information $\tilde{Z}i$ serves as the initial input, which is applied iteratively for $N$ steps to update the $D^l_{ic}$ based on its local environment. As shown in Fig. \ref{fig_1}g, each interaction block consists of several submodules. The self-interaction block performs intra-channel mixing across the descriptor components $D_{ic}^l$ via a linear transformation:
\begin{equation}
    D_{ic}^l \leftarrow \sum_{c'}W_{cc'}^l D_{ic'}^l
\end{equation}
where $W_{cc'}^l$ is a learnable weight tensor. Following self-interaction, the descriptors are passed through a convolution block, which aggregates information from neighboring atoms using a learnable function of relative positions and their descriptors. This operation is formally expressed as:
\begin{equation}
    \mathcal{L}^{l_o}_{ic}\left(\bm{R}_i,D_{ic}^{l_q}\right) = \sum_{j\in \mathcal{N}(i)}F_{c}^{l_pl_q}(\bm{R}_{ij})\otimes_{l_p,l_q \to l_o}D_{jc}^{l_q}
\end{equation}
where $\mathcal{N}(i)$ denotes the neighbors of atom $i$, $F_{c}^{ll'}$ is a learnable kernel function for $(l,l')$ pairs depending on the relative position $\bm{R}_{ij}$. The kernel is constrained to be $SO(3)$-equivariant and takes the form:
\begin{equation}
    F_{c}^{ll'}(\bm{r})=R_c^{ll'}\left(B(r)\cdot \mathrm{smooth}(r)\right)Y^{l}(\hat{\bm{r}})
\end{equation}
in order to be equivariant to $SO(3)$, where $r=|\bm{r}|$ and $\hat{\bm{r}}$ is the corresponding unit vector. $Y^l$ denotes the spherical harmonics, $B(r)$ is an edge embedding function that projects the radial distance onto a set of basis functions (e.g., radial Bessel or Gaussian functions), and smooth$(r)$ is a radial smooth function that gradually decreases to zero at the cutoff radius. $R_c^{ll'}$ are multi-layer perceptrons (MLP) for each $(l,l')$ pair. Since tensor product operations can yield multiple outputs $\mathcal{L}^{l_o}_{ic_o}$ for the same angular momentum order $l_o$, a concatenation step is applied to merge all such outputs. To ensure tractability, a maximum angular momentum order $L$ is imposed, limiting the representation to finite $l$ values and improving computational efficiency. Each layer concludes with a residual connection followed by a non-linear activation, specifically the SiLU function, to introduce non-linearity into the learned features.

\subsection{Ablation studies}

We conducted comprehensive ablation studies on the EAC-Net model using two representative material systems: $\mathrm{H_2O}$ molecular system and Al-Mg alloy (Al:Mg = 1:1, 54 atoms each). All models shared unified hyperparameters: $L=3$, feature length of 36, four atom convolution layers, two edge update layers, and approximately 1 million parameters. The learning rate followed a multi-stage exponential decay schedule from $3\times10^{-3}$ to $2\times10^{-4}$ over 110,000 training steps. The atomic cutoff radius was set to 4.0 \AA~and the atom-probe cutoff radius to 6.0 \AA.

The $\mathrm{H_2O}$ system training set contained 6 frames (3$\times$5,000 randomly sampled points per frame using three sampling strategies: pure random, density-weighted, and gradient-weighted), with 4 frames for testing. The Al-Mg alloy training set included 12 frames ($\mathrm{216^3}$ grid points each) with 2 frames for testing. All reported results are test set $\varepsilon_{\mathrm{mae}}$ from single training runs.

\subsubsection{Weight net architecture}

To systematically evaluate the impact of different weighting strategies in our model, we conducted an ablation study by testing several alternative formulations of the $w$. Specifically, we tested the following variants:

\begin{enumerate}[(i)]
    \item \textbf{Identity (I)}: This baseline method uses only an identity mapping.
    
    \item \textbf{Smooth Function Only (S)}: This variant computes the weights purely based on the distance-based smooth function:
    \[
    w(G_{i}, r_{i}) = \mathrm{smooth}(r_{i})
    \]

    \item \textbf{MLP+S}: The output from the MLP and the smooth function are summed before applying softmax:
    \[
    w(G_{i}, r_{i}) =\mathrm{softmax} \left( \mathrm{MLP}\left(G_{i}\mid\sum_{j\in \mathcal{N}\left( \bm{r} \right)}{G_{j}} \right) + \mathrm{smooth}\left( r_{i} \right) \right) 
    \]

    \item \textbf{MLP$\times$S}: The output from the MLP and the smooth function are multiplied before applying softmax:
    \[
    w(G_{i}, r_{i}) =\mathrm{softmax} \left( \mathrm{MLP}\left(G_{i}\mid\sum_{j\in \mathcal{N}\left( \bm r \right)}{G_{j}} \right) \times \mathrm{smooth}\left( r_{i} \right) \right) 
    \]

    \item \textbf{Softmax$\times$S}: The smooth function is applied after softmax over the MLP output:
    \[
    w(G_{i}, r_{i}) =\mathrm{softmax} \left( \mathrm{MLP}\left(G_{i}\mid\sum_{j\in \mathcal{N}\left( \bm r \right)}{G_{j}} \right) \right) \times \mathrm{smooth}\left( r_{i} \right)
    \]
\end{enumerate}

Table \ref{tab:ablation-weights} summarizes the performance of each method across $\mathrm{H_2O}$ and Al-Mg systems. For the $\mathrm{H_2O}$ system, the MLP$\times$S strategy performed optimally, achieving approximately 45-fold improvement over the baseline. For the Al-Mg alloy, the MLP+S was most effective, providing about 3.8-fold improvement over the baseline.

\begin{table}[]
\centering
\caption{$\varepsilon_{\mathrm{mae}}$ of different weight function variants in $\mathrm{H_2O}$ and Al-Mg.}
\begin{tabular}{@{\extracolsep\fill}lcc}
\toprule
$w$ & $\mathrm{H_2O}$ (\%) & Al-Mg (\%) \\
\midrule
I                   & 9.47          & 1.54 \\
S                   & 1.22          & 0.72 \\
MLP+S               & 0.25          & \textbf{0.41} \\
MLP$\times$S        & \textbf{0.21} & 1.03 \\
Softmax$\times$S    & 0.28          & 3.22 \\
\botrule
\end{tabular}
\label{tab:ablation-weights}
\end{table}

\subsubsection{Feature concatenation strategies}

To explore the effect of residual connection concepts in atom-probe features, we tested different attribute concatenation approaches, as listed in Table \ref{tab:ablation-cat}. Feature concatenation strategies showed different patterns across the two systems. For the $\mathrm{H_2O}$ system, the baseline approach without additional concatenation performed best, with supplementary features potentially introducing noise. For the bulk Al-Mg alloy system, full feature concatenation outperformed other methods, achieving approximately 1.7-fold improvement over the baseline, indicating that complex bulk systems can benefit from comprehensive feature representations that include geometric and chemical information.

\begin{table}[]
\centering
\caption{$\varepsilon_{\mathrm{mae}}$ of different feature concatenation strategies in $\mathrm{H_2O}$ and Al-Mg.}
\begin{tabular}{@{\extracolsep\fill}lcc}
\toprule
$\mathcal{S}$ & $\mathrm{H_2O}$ (\%) & Al-Mg (\%) \\
\midrule
$\{ \}$                             & \textbf{0.21}   & 0.41 \\
$\{r_{i}\}$                   & 0.75            & 0.61 \\
$\{B(r_{i})\}$                & 0.26            & 0.73 \\
$\{Z_{i}\}$                         & \textbf{0.21}     & 0.36 \\
$\{r_{i}$, $B(r_{i})$, $Z_{i}\}$   & 0.29   & \textbf{0.24} \\
\botrule
\end{tabular}
\label{tab:ablation-cat}
\end{table}

\subsubsection{Type embedding strategies}

We compared three atomic representation approaches listed in Table \ref{tab:ablation-type}. For the $\mathrm{H_2O}$ system, all three encoding methods performed comparably, with differences within the range of training randomness. For the Al-Mg alloy, the hybrid encoding approach performed best.

\begin{table}[]
\centering
\caption{$\varepsilon_{\mathrm{mae}}$ of different type embedding strategies in $\mathrm{H_2O}$ and Al-Mg.}
\begin{tabular}{@{\extracolsep\fill}lcc}
\toprule
Method & $\mathrm{H_2O}$ (\%) & Al-Mg (\%) \\
\midrule
Shell-filling             & \textbf{0.21}   & 0.24 \\
One-hot                   & 0.23            & 4.55 \\
Shell-filling\texttt{|}One-hot   & 0.25     & \textbf{0.24} \\
\botrule
\end{tabular}
\label{tab:ablation-type}
\end{table}

\subsection{Training details}

The hyperparameter settings described below correspond to the performance evaluation. Training was performed using the Adam optimizer with an initial learning rate of 0.003. Learning rate schedules followed a multi-stage exponential decay scheme, typically reducing the rate to around 10$^{-7}$, depending on the task. Models were trained for 150,000 steps with a batch size of 50. For Al-Mg, the total training steps were set to 250,000 with an end learning rate of 10$^{-8}$. Radial cutoffs of 4.0 \AA~(for atom features) and 6.0 \AA~(for atom-grid coupling) were used to define local environments. The weight net architectures were MLP+S. For the spin density task in Fe, we trained for 200,000 steps with a final learning rate of 10$^{-5}$. The number of convolution layers, edge updatings, $L$, and feature dimensionality were adjusted per system to account for structural and chemical complexity. Detailed hyperparameter settings for each model system are provided in the Supplementary Information.

The EAC-mp model was trained using 2$\times$ NVIDIA RTX A6000 GPUs (48\,GB each). The model architecture employed an angular momentum cutoff of $L=5$, a feature dimensionality of 36, four atom interaction convolutional layers, and two edge updating layers. The weight net adopts the MLP+S architecture. The total number of trainable parameters was approximately 3.08 million.

The training dataset comprised 48,183 configurations, each associated with 5000 sampling points extracted from the charge density. Training was performed over 600,000 steps. The learning rate decayed exponentially from $1.0 \times 10^{-3}$ to $2.3 \times 10^{-5}$. The atom and grid cutoffs were set to 4.0\,\AA\ and 6.0\,\AA, respectively. The batch size varied across different stages of training, ranging from 400 to 750 samples per step.

\backmatter

\subsection*{Data availability}
Materials Project data were collected in August 2025, the task identifiers are included in our provided repository. The preprocessed dataset used for training the EAC-mp model is available at \url{https://zenodo.org/records/16990467}. Datasets for the task-specific models can be obtained from the authors upon reasonable request. The pretrained model weight of EAC-mp is available at \url{https://www.aissquare.com/models/detail?pageType=models&name=EAC-mp-l5-3M&id=368}.

\subsection*{Code availability}
Our model implementation is available at \url{https://github.com/qin2xue3jian4/EAC-Net}.

\subsection*{Acknowledgements}
This work was supported by the National Key R\&D Program of China (Grants No.2021YFA0718900), the National Natural Science Foundation of China (Grants No.12374096 and No.92477114). This research is also supported by the Jiangsu Funding Program for Excellent Postdoctoral Talent.

\subsection*{Author contributions}
X.Q. conceived the EAC-Net model, implemented the software, created the data set for EAC-mp, and conducted some software experiments under the guidance of T.L. and Z.Z., T.L. performed data analysis, drafted the manuscript, and contributed to the development of the model, Z.Z. supervised the project from conception to design of experiments, implementation, theory, and analysis of data. All authors contributed to writing the manuscript.

\subsection*{Competing interests}
The authors declare no competing interests.

\subsection*{Additional information}

\textbf{Supplementary information} The online version contains supplementary material.

\textbf{Correspondence} and requests for materials should be addressed to \href{zczhong@ustc.edu.cn}{Z.Z.}.


\bibliography{sn-bibliography}


\begin{thebibliography}{31}
\ifx \bisbn   \undefined \def \bisbn  #1{ISBN #1}\fi
\ifx \binits  \undefined \def \binits#1{#1}\fi
\ifx \bauthor  \undefined \def \bauthor#1{#1}\fi
\ifx \batitle  \undefined \def \batitle#1{#1}\fi
\ifx \bjtitle  \undefined \def \bjtitle#1{#1}\fi
\ifx \bvolume  \undefined \def \bvolume#1{\textbf{#1}}\fi
\ifx \byear  \undefined \def \byear#1{#1}\fi
\ifx \bissue  \undefined \def \bissue#1{#1}\fi
\ifx \bfpage  \undefined \def \bfpage#1{#1}\fi
\ifx \blpage  \undefined \def \blpage #1{#1}\fi
\ifx \burl  \undefined \def \burl#1{\textsf{#1}}\fi
\ifx \doiurl  \undefined \def \doiurl#1{\url{https://doi.org/#1}}\fi
\ifx \betal  \undefined \def \betal{\textit{et al.}}\fi
\ifx \binstitute  \undefined \def \binstitute#1{#1}\fi
\ifx \binstitutionaled  \undefined \def \binstitutionaled#1{#1}\fi
\ifx \bctitle  \undefined \def \bctitle#1{#1}\fi
\ifx \beditor  \undefined \def \beditor#1{#1}\fi
\ifx \bpublisher  \undefined \def \bpublisher#1{#1}\fi
\ifx \bbtitle  \undefined \def \bbtitle#1{#1}\fi
\ifx \bedition  \undefined \def \bedition#1{#1}\fi
\ifx \bseriesno  \undefined \def \bseriesno#1{#1}\fi
\ifx \blocation  \undefined \def \blocation#1{#1}\fi
\ifx \bsertitle  \undefined \def \bsertitle#1{#1}\fi
\ifx \bsnm \undefined \def \bsnm#1{#1}\fi
\ifx \bsuffix \undefined \def \bsuffix#1{#1}\fi
\ifx \bparticle \undefined \def \bparticle#1{#1}\fi
\ifx \barticle \undefined \def \barticle#1{#1}\fi
\bibcommenthead
\ifx \bconfdate \undefined \def \bconfdate #1{#1}\fi
\ifx \botherref \undefined \def \botherref #1{#1}\fi
\ifx \url \undefined \def \url#1{\textsf{#1}}\fi
\ifx \bchapter \undefined \def \bchapter#1{#1}\fi
\ifx \bbook \undefined \def \bbook#1{#1}\fi
\ifx \bcomment \undefined \def \bcomment#1{#1}\fi
\ifx \oauthor \undefined \def \oauthor#1{#1}\fi
\ifx \citeauthoryear \undefined \def \citeauthoryear#1{#1}\fi
\ifx \endbibitem  \undefined \def \endbibitem {}\fi
\ifx \bconflocation  \undefined \def \bconflocation#1{#1}\fi
\ifx \arxivurl  \undefined \def \arxivurl#1{\textsf{#1}}\fi
\csname PreBibitemsHook\endcsname

\bibitem[\protect\citeauthoryear{Kohn and Sham}{1965}]{PhysRev.140.A1133}
\begin{barticle}
\bauthor{\bsnm{Kohn}, \binits{W.}},
\bauthor{\bsnm{Sham}, \binits{L.J.}}:
\batitle{Self-consistent equations including exchange and correlation effects}.
\bjtitle{Physical Review}
\bvolume{140}(\bissue{4A}),
\bfpage{1133}--\blpage{1138}
(\byear{1965})
\doiurl{10.1103/PhysRev.140.A1133}
\end{barticle}
\endbibitem

\bibitem[\protect\citeauthoryear{Hohenberg and Kohn}{1964}]{PhysRev.136.B864}
\begin{barticle}
\bauthor{\bsnm{Hohenberg}, \binits{P.}},
\bauthor{\bsnm{Kohn}, \binits{W.}}:
\batitle{Inhomogeneous electron gas}.
\bjtitle{Physical Review}
\bvolume{136}(\bissue{3B}),
\bfpage{864}--\blpage{871}
(\byear{1964})
\doiurl{10.1103/PhysRev.136.B864}
\end{barticle}
\endbibitem

\bibitem[\protect\citeauthoryear{Batatia et~al.}{2024}]{batatia2024foundationmodelatomisticmaterials}
\begin{botherref}
\oauthor{\bsnm{Batatia}, \binits{I.}},
\oauthor{\bsnm{Benner}, \binits{P.}},
\oauthor{\bsnm{Chiang}, \binits{Y.}},
\oauthor{\bsnm{Elena}, \binits{A.M.}},
\oauthor{\bsnm{Kovács}, \binits{D.P.}},
\oauthor{\bsnm{Riebesell}, \binits{J.}},
\oauthor{\bsnm{Advincula}, \binits{X.R.}},
\oauthor{\bsnm{Asta}, \binits{M.}},
\oauthor{\bsnm{Avaylon}, \binits{M.}},
\oauthor{\bsnm{Baldwin}, \binits{W.J.}},
\oauthor{\bsnm{Berger}, \binits{F.}},
\oauthor{\bsnm{Bernstein}, \binits{N.}},
\oauthor{\bsnm{Bhowmik}, \binits{A.}},
\oauthor{\bsnm{Blau}, \binits{S.M.}},
\oauthor{\bsnm{Cărare}, \binits{V.}},
\oauthor{\bsnm{Darby}, \binits{J.P.}},
\oauthor{\bsnm{De}, \binits{S.}},
\oauthor{\bsnm{Pia}, \binits{F.D.}},
\oauthor{\bsnm{Deringer}, \binits{V.L.}},
\oauthor{\bsnm{Elijošius}, \binits{R.}},
\oauthor{\bsnm{El-Machachi}, \binits{Z.}},
\oauthor{\bsnm{Falcioni}, \binits{F.}},
\oauthor{\bsnm{Fako}, \binits{E.}},
\oauthor{\bsnm{Ferrari}, \binits{A.C.}},
\oauthor{\bsnm{Genreith-Schriever}, \binits{A.}},
\oauthor{\bsnm{George}, \binits{J.}},
\oauthor{\bsnm{Goodall}, \binits{R.E.A.}},
\oauthor{\bsnm{Grey}, \binits{C.P.}},
\oauthor{\bsnm{Grigorev}, \binits{P.}},
\oauthor{\bsnm{Han}, \binits{S.}},
\oauthor{\bsnm{Handley}, \binits{W.}},
\oauthor{\bsnm{Heenen}, \binits{H.H.}},
\oauthor{\bsnm{Hermansson}, \binits{K.}},
\oauthor{\bsnm{Holm}, \binits{C.}},
\oauthor{\bsnm{Jaafar}, \binits{J.}},
\oauthor{\bsnm{Hofmann}, \binits{S.}},
\oauthor{\bsnm{Jakob}, \binits{K.S.}},
\oauthor{\bsnm{Jung}, \binits{H.}},
\oauthor{\bsnm{Kapil}, \binits{V.}},
\oauthor{\bsnm{Kaplan}, \binits{A.D.}},
\oauthor{\bsnm{Karimitari}, \binits{N.}},
\oauthor{\bsnm{Kermode}, \binits{J.R.}},
\oauthor{\bsnm{Kroupa}, \binits{N.}},
\oauthor{\bsnm{Kullgren}, \binits{J.}},
\oauthor{\bsnm{Kuner}, \binits{M.C.}},
\oauthor{\bsnm{Kuryla}, \binits{D.}},
\oauthor{\bsnm{Liepuoniute}, \binits{G.}},
\oauthor{\bsnm{Margraf}, \binits{J.T.}},
\oauthor{\bsnm{Magdău}, \binits{I.-B.}},
\oauthor{\bsnm{Michaelides}, \binits{A.}},
\oauthor{\bsnm{Moore}, \binits{J.H.}},
\oauthor{\bsnm{Naik}, \binits{A.A.}},
\oauthor{\bsnm{Niblett}, \binits{S.P.}},
\oauthor{\bsnm{Norwood}, \binits{S.W.}},
\oauthor{\bsnm{O'Neill}, \binits{N.}},
\oauthor{\bsnm{Ortner}, \binits{C.}},
\oauthor{\bsnm{Persson}, \binits{K.A.}},
\oauthor{\bsnm{Reuter}, \binits{K.}},
\oauthor{\bsnm{Rosen}, \binits{A.S.}},
\oauthor{\bsnm{Schaaf}, \binits{L.L.}},
\oauthor{\bsnm{Schran}, \binits{C.}},
\oauthor{\bsnm{Shi}, \binits{B.X.}},
\oauthor{\bsnm{Sivonxay}, \binits{E.}},
\oauthor{\bsnm{Stenczel}, \binits{T.K.}},
\oauthor{\bsnm{Svahn}, \binits{V.}},
\oauthor{\bsnm{Sutton}, \binits{C.}},
\oauthor{\bsnm{Swinburne}, \binits{T.D.}},
\oauthor{\bsnm{Tilly}, \binits{J.}},
\oauthor{\bsnm{Oord}, \binits{C.}},
\oauthor{\bsnm{Varga-Umbrich}, \binits{E.}},
\oauthor{\bsnm{Vegge}, \binits{T.}},
\oauthor{\bsnm{Vondrák}, \binits{M.}},
\oauthor{\bsnm{Wang}, \binits{Y.}},
\oauthor{\bsnm{Witt}, \binits{W.C.}},
\oauthor{\bsnm{Zills}, \binits{F.}},
\oauthor{\bsnm{Csányi}, \binits{G.}}:
A foundation model for atomistic materials chemistry
(2024).
\url{https://arxiv.org/abs/2401.00096}
\end{botherref}
\endbibitem

\bibitem[\protect\citeauthoryear{Yang et~al.}{2024}]{yang2024mattersimdeeplearningatomistic}
\begin{botherref}
\oauthor{\bsnm{Yang}, \binits{H.}},
\oauthor{\bsnm{Hu}, \binits{C.}},
\oauthor{\bsnm{Zhou}, \binits{Y.}},
\oauthor{\bsnm{Liu}, \binits{X.}},
\oauthor{\bsnm{Shi}, \binits{Y.}},
\oauthor{\bsnm{Li}, \binits{J.}},
\oauthor{\bsnm{Li}, \binits{G.}},
\oauthor{\bsnm{Chen}, \binits{Z.}},
\oauthor{\bsnm{Chen}, \binits{S.}},
\oauthor{\bsnm{Zeni}, \binits{C.}},
\oauthor{\bsnm{Horton}, \binits{M.}},
\oauthor{\bsnm{Pinsler}, \binits{R.}},
\oauthor{\bsnm{Fowler}, \binits{A.}},
\oauthor{\bsnm{Zügner}, \binits{D.}},
\oauthor{\bsnm{Xie}, \binits{T.}},
\oauthor{\bsnm{Smith}, \binits{J.}},
\oauthor{\bsnm{Sun}, \binits{L.}},
\oauthor{\bsnm{Wang}, \binits{Q.}},
\oauthor{\bsnm{Kong}, \binits{L.}},
\oauthor{\bsnm{Liu}, \binits{C.}},
\oauthor{\bsnm{Hao}, \binits{H.}},
\oauthor{\bsnm{Lu}, \binits{Z.}}:
MatterSim: A Deep Learning Atomistic Model Across Elements, Temperatures and Pressures
(2024).
\url{https://arxiv.org/abs/2405.04967}
\end{botherref}
\endbibitem

\bibitem[\protect\citeauthoryear{Zeng et~al.}{2025}]{zengDeePMDkitV3MultipleBackend2025}
\begin{barticle}
\bauthor{\bsnm{Zeng}, \binits{J.}},
\bauthor{\bsnm{Zhang}, \binits{D.}},
\bauthor{\bsnm{Peng}, \binits{A.}},
\bauthor{\bsnm{Zhang}, \binits{X.}},
\bauthor{\bsnm{He}, \binits{S.}},
\bauthor{\bsnm{Wang}, \binits{Y.}},
\bauthor{\bsnm{Liu}, \binits{X.}},
\bauthor{\bsnm{Bi}, \binits{H.}},
\bauthor{\bsnm{Li}, \binits{Y.}},
\bauthor{\bsnm{Cai}, \binits{C.}},
\bauthor{\bsnm{Zhang}, \binits{C.}},
\bauthor{\bsnm{Du}, \binits{Y.}},
\bauthor{\bsnm{Zhu}, \binits{J.-X.}},
\bauthor{\bsnm{Mo}, \binits{P.}},
\bauthor{\bsnm{Huang}, \binits{Z.}},
\bauthor{\bsnm{Zeng}, \binits{Q.}},
\bauthor{\bsnm{Shi}, \binits{S.}},
\bauthor{\bsnm{Qin}, \binits{X.}},
\bauthor{\bsnm{Yu}, \binits{Z.}},
\bauthor{\bsnm{Luo}, \binits{C.}},
\bauthor{\bsnm{Ding}, \binits{Y.}},
\bauthor{\bsnm{Liu}, \binits{Y.-P.}},
\bauthor{\bsnm{Shi}, \binits{R.}},
\bauthor{\bsnm{Wang}, \binits{Z.}},
\bauthor{\bsnm{Bore}, \binits{S.L.}},
\bauthor{\bsnm{Chang}, \binits{J.}},
\bauthor{\bsnm{Deng}, \binits{Z.}},
\bauthor{\bsnm{Ding}, \binits{Z.}},
\bauthor{\bsnm{Han}, \binits{S.}},
\bauthor{\bsnm{Jiang}, \binits{W.}},
\bauthor{\bsnm{Ke}, \binits{G.}},
\bauthor{\bsnm{Liu}, \binits{Z.}},
\bauthor{\bsnm{Lu}, \binits{D.}},
\bauthor{\bsnm{Muraoka}, \binits{K.}},
\bauthor{\bsnm{Oliaei}, \binits{H.}},
\bauthor{\bsnm{Singh}, \binits{A.K.}},
\bauthor{\bsnm{Que}, \binits{H.}},
\bauthor{\bsnm{Xu}, \binits{W.}},
\bauthor{\bsnm{Xu}, \binits{Z.}},
\bauthor{\bsnm{Zhuang}, \binits{Y.-B.}},
\bauthor{\bsnm{Dai}, \binits{J.}},
\bauthor{\bsnm{Giese}, \binits{T.J.}},
\bauthor{\bsnm{Jia}, \binits{W.}},
\bauthor{\bsnm{Xu}, \binits{B.}},
\bauthor{\bsnm{York}, \binits{D.M.}},
\bauthor{\bsnm{Zhang}, \binits{L.}},
\bauthor{\bsnm{Wang}, \binits{H.}}:
\batitle{{{DeePMD-kit}} v3: {{A Multiple-Backend Framework}} for {{Machine Learning Potentials}}}.
\bjtitle{Journal of Chemical Theory and Computation}
\bvolume{21}(\bissue{9}),
\bfpage{4375}--\blpage{4385}
(\byear{2025})
\doiurl{10.1021/acs.jctc.5c00340}
\end{barticle}
\endbibitem

\bibitem[\protect\citeauthoryear{Fu et~al.}{2025}]{fu2025learningsmoothexpressiveinteratomic}
\begin{botherref}
\oauthor{\bsnm{Fu}, \binits{X.}},
\oauthor{\bsnm{Wood}, \binits{B.M.}},
\oauthor{\bsnm{Barroso-Luque}, \binits{L.}},
\oauthor{\bsnm{Levine}, \binits{D.S.}},
\oauthor{\bsnm{Gao}, \binits{M.}},
\oauthor{\bsnm{Dzamba}, \binits{M.}},
\oauthor{\bsnm{Zitnick}, \binits{C.L.}}:
Learning Smooth and Expressive Interatomic Potentials for Physical Property Prediction
(2025).
\url{https://arxiv.org/abs/2502.12147}
\end{botherref}
\endbibitem

\bibitem[\protect\citeauthoryear{Kim et~al.}{2025}]{kimDataEfficientMultifidelityTraining2025}
\begin{barticle}
\bauthor{\bsnm{Kim}, \binits{J.}},
\bauthor{\bsnm{Kim}, \binits{J.}},
\bauthor{\bsnm{Kim}, \binits{J.}},
\bauthor{\bsnm{Lee}, \binits{J.}},
\bauthor{\bsnm{Park}, \binits{Y.}},
\bauthor{\bsnm{Kang}, \binits{Y.}},
\bauthor{\bsnm{Han}, \binits{S.}}:
\batitle{Data-{{Efficient Multifidelity Training}} for {{High-Fidelity Machine Learning Interatomic Potentials}}}.
\bjtitle{Journal of the American Chemical Society}
\bvolume{147}(\bissue{1}),
\bfpage{1042}--\blpage{1054}
(\byear{2025})
\doiurl{10.1021/jacs.4c14455}
\end{barticle}
\endbibitem

\bibitem[\protect\citeauthoryear{Koker et~al.}{2024}]{kokerHigherorderEquivariantNeural2024a}
\begin{barticle}
\bauthor{\bsnm{Koker}, \binits{T.}},
\bauthor{\bsnm{Quigley}, \binits{K.}},
\bauthor{\bsnm{Taw}, \binits{E.}},
\bauthor{\bsnm{Tibbetts}, \binits{K.}},
\bauthor{\bsnm{Li}, \binits{L.}}:
\batitle{Higher-order equivariant neural networks for charge density prediction in materials}.
\bjtitle{npj Computational Materials}
\bvolume{10}(\bissue{1}),
\bfpage{161}
(\byear{2024})
\doiurl{10.1038/s41524-024-01343-1}
\end{barticle}
\endbibitem

\bibitem[\protect\citeauthoryear{Lv et~al.}{2023}]{PhysRevB.108.235159}
\begin{barticle}
\bauthor{\bsnm{Lv}, \binits{T.}},
\bauthor{\bsnm{Zhong}, \binits{Z.}},
\bauthor{\bsnm{Liang}, \binits{Y.}},
\bauthor{\bsnm{Li}, \binits{F.}},
\bauthor{\bsnm{Huang}, \binits{J.}},
\bauthor{\bsnm{Zheng}, \binits{R.}}:
\batitle{Deep {{Charge}}: {{Deep}} learning model of electron density from a one-shot density functional theory calculation}.
\bjtitle{Physical Review B}
\bvolume{108}(\bissue{23}),
\bfpage{235159}
(\byear{2023})
\doiurl{10.1103/PhysRevB.108.235159}
\end{barticle}
\endbibitem

\bibitem[\protect\citeauthoryear{J{\o}rgensen and Bhowmik}{2022}]{jorgensenEquivariantGraphNeural2022}
\begin{barticle}
\bauthor{\bsnm{J{\o}rgensen}, \binits{P.B.}},
\bauthor{\bsnm{Bhowmik}, \binits{A.}}:
\batitle{Equivariant graph neural networks for fast electron density estimation of molecules, liquids, and solids}.
\bjtitle{npj Computational Materials}
\bvolume{8}(\bissue{1}),
\bfpage{183}
(\byear{2022})
\doiurl{10.1038/s41524-022-00863-y}
\end{barticle}
\endbibitem

\bibitem[\protect\citeauthoryear{{Zepeda-N{\'u}{\~n}ez} et~al.}{2021}]{ZEPEDANUNEZ2021110523}
\begin{barticle}
\bauthor{\bsnm{{Zepeda-N{\'u}{\~n}ez}}, \binits{L.}},
\bauthor{\bsnm{Chen}, \binits{Y.}},
\bauthor{\bsnm{Zhang}, \binits{J.}},
\bauthor{\bsnm{Jia}, \binits{W.}},
\bauthor{\bsnm{Zhang}, \binits{L.}},
\bauthor{\bsnm{Lin}, \binits{L.}}:
\batitle{Deep {{Density}}: {{Circumventing}} the {{Kohn-Sham}} equations via symmetry preserving neural networks}.
\bjtitle{Journal of Computational Physics}
\bvolume{443},
\bfpage{110523}
(\byear{2021})
\doiurl{10.1016/j.jcp.2021.110523}
\end{barticle}
\endbibitem

\bibitem[\protect\citeauthoryear{Focassio et~al.}{2024}]{PhysRevB.110.184106}
\begin{barticle}
\bauthor{\bsnm{Focassio}, \binits{B.}},
\bauthor{\bsnm{Domina}, \binits{M.}},
\bauthor{\bsnm{Patil}, \binits{U.}},
\bauthor{\bsnm{Fazzio}, \binits{A.}},
\bauthor{\bsnm{Sanvito}, \binits{S.}}:
\batitle{Covariant {{Jacobi-Legendre}} expansion for total energy calculations within the projector augmented wave formalism}.
\bjtitle{Physical Review B}
\bvolume{110}(\bissue{18}),
\bfpage{184106}
(\byear{2024})
\doiurl{10.1103/PhysRevB.110.184106}
\end{barticle}
\endbibitem

\bibitem[\protect\citeauthoryear{Chandrasekaran et~al.}{2019}]{chandrasekaranSolvingElectronicStructure2019a}
\begin{barticle}
\bauthor{\bsnm{Chandrasekaran}, \binits{A.}},
\bauthor{\bsnm{Kamal}, \binits{D.}},
\bauthor{\bsnm{Batra}, \binits{R.}},
\bauthor{\bsnm{Kim}, \binits{C.}},
\bauthor{\bsnm{Chen}, \binits{L.}},
\bauthor{\bsnm{Ramprasad}, \binits{R.}}:
\batitle{Solving the electronic structure problem with machine learning}.
\bjtitle{npj Computational Materials}
\bvolume{5}(\bissue{1}),
\bfpage{22}
(\byear{2019})
\doiurl{10.1038/s41524-019-0162-7}
\end{barticle}
\endbibitem

\bibitem[\protect\citeauthoryear{Focassio et~al.}{2023}]{focassioLinearJacobiLegendreExpansion2023a}
\begin{barticle}
\bauthor{\bsnm{Focassio}, \binits{B.}},
\bauthor{\bsnm{Domina}, \binits{M.}},
\bauthor{\bsnm{Patil}, \binits{U.}},
\bauthor{\bsnm{Fazzio}, \binits{A.}},
\bauthor{\bsnm{Sanvito}, \binits{S.}}:
\batitle{Linear {{Jacobi-Legendre}} expansion of the charge density for machine learning-accelerated electronic structure calculations}.
\bjtitle{npj Computational Materials}
\bvolume{9}(\bissue{1}),
\bfpage{87}
(\byear{2023})
\doiurl{10.1038/s41524-023-01053-0}
\end{barticle}
\endbibitem

\bibitem[\protect\citeauthoryear{Lewis et~al.}{2021}]{lewisLearningElectronDensities2021}
\begin{barticle}
\bauthor{\bsnm{Lewis}, \binits{A.M.}},
\bauthor{\bsnm{Grisafi}, \binits{A.}},
\bauthor{\bsnm{Ceriotti}, \binits{M.}},
\bauthor{\bsnm{Rossi}, \binits{M.}}:
\batitle{Learning {{Electron Densities}} in the {{Condensed Phase}}}.
\bjtitle{Journal of Chemical Theory and Computation}
\bvolume{17}(\bissue{11}),
\bfpage{7203}--\blpage{7214}
(\byear{2021})
\doiurl{10.1021/acs.jctc.1c00576}
\end{barticle}
\endbibitem

\bibitem[\protect\citeauthoryear{Grisafi et~al.}{2019}]{grisafiTransferableMachineLearningModel2019}
\begin{barticle}
\bauthor{\bsnm{Grisafi}, \binits{A.}},
\bauthor{\bsnm{Fabrizio}, \binits{A.}},
\bauthor{\bsnm{Meyer}, \binits{B.}},
\bauthor{\bsnm{Wilkins}, \binits{D.M.}},
\bauthor{\bsnm{Corminboeuf}, \binits{C.}},
\bauthor{\bsnm{Ceriotti}, \binits{M.}}:
\batitle{Transferable {{Machine-Learning Model}} of the {{Electron Density}}}.
\bjtitle{ACS Central Science}
\bvolume{5}(\bissue{1}),
\bfpage{57}--\blpage{64}
(\byear{2019})
\doiurl{10.1021/acscentsci.8b00551}
\end{barticle}
\endbibitem

\bibitem[\protect\citeauthoryear{Fu et~al.}{2025}]{NEURIPS2024_12c118ef}
\begin{bchapter}
\bauthor{\bsnm{Fu}, \binits{X.}},
\bauthor{\bsnm{Rosen}, \binits{A.}},
\bauthor{\bsnm{Bystrom}, \binits{K.}},
\bauthor{\bsnm{Wang}, \binits{R.}},
\bauthor{\bsnm{Musaelian}, \binits{A.}},
\bauthor{\bsnm{Kozinsky}, \binits{B.}},
\bauthor{\bsnm{Smidt}, \binits{T.}},
\bauthor{\bsnm{Jaakkola}, \binits{T.}}:
\bctitle{A recipe for charge density prediction}.
In: \bbtitle{Proceedings of the 38th International Conference on Neural Information Processing Systems}.
\bsertitle{NIPS '24}.
\bpublisher{Curran Associates Inc.},
\blocation{Red Hook, NY, USA}
(\byear{2025})
\end{bchapter}
\endbibitem

\bibitem[\protect\citeauthoryear{Fabrizio et~al.}{2019}]{C9SC02696G}
\begin{barticle}
\bauthor{\bsnm{Fabrizio}, \binits{A.}},
\bauthor{\bsnm{Grisafi}, \binits{A.}},
\bauthor{\bsnm{Meyer}, \binits{B.}},
\bauthor{\bsnm{Ceriotti}, \binits{M.}},
\bauthor{\bsnm{Corminboeuf}, \binits{C.}}:
\batitle{Electron density learning of non-covalent systems}.
\bjtitle{Chemical Science}
\bvolume{10}(\bissue{41}),
\bfpage{9424}--\blpage{9432}
(\byear{2019})
\doiurl{10.1039/C9SC02696G}
\end{barticle}
\endbibitem

\bibitem[\protect\citeauthoryear{Munro et~al.}{2020}]{Munro2020}
\begin{botherref}
\oauthor{\bsnm{Munro}, \binits{J.M.}},
\oauthor{\bsnm{Latimer}, \binits{K.}},
\oauthor{\bsnm{Horton}, \binits{M.K.}},
\oauthor{\bsnm{Dwaraknath}, \binits{S.}},
\oauthor{\bsnm{Persson}, \binits{K.A.}}:
An improved symmetry-based approach to reciprocal space path selection in band structure calculations.
npj Computational Materials
\textbf{6}(1)
(2020)
\doiurl{10.1038/s41524-020-00383-7}
\end{botherref}
\endbibitem

\bibitem[\protect\citeauthoryear{Merchant et~al.}{2023}]{Merchant2023}
\begin{barticle}
\bauthor{\bsnm{Merchant}, \binits{A.}},
\bauthor{\bsnm{Batzner}, \binits{S.}},
\bauthor{\bsnm{Schoenholz}, \binits{S.S.}},
\bauthor{\bsnm{Aykol}, \binits{M.}},
\bauthor{\bsnm{Cheon}, \binits{G.}},
\bauthor{\bsnm{Cubuk}, \binits{E.D.}}:
\batitle{Scaling deep learning for materials discovery}.
\bjtitle{Nature}
\bvolume{624}(\bissue{7990}),
\bfpage{80}--\blpage{85}
(\byear{2023})
\doiurl{10.1038/s41586-023-06735-9}
\end{barticle}
\endbibitem

\bibitem[\protect\citeauthoryear{Geiger and Smidt}{2022}]{geiger2022e3nneuclideanneuralnetworks}
\begin{botherref}
\oauthor{\bsnm{Geiger}, \binits{M.}},
\oauthor{\bsnm{Smidt}, \binits{T.}}:
e3nn: Euclidean Neural Networks
(2022).
\url{https://arxiv.org/abs/2207.09453}
\end{botherref}
\endbibitem

\bibitem[\protect\citeauthoryear{Batzner et~al.}{2022}]{batznerE3equivariantGraphNeural2022}
\begin{barticle}
\bauthor{\bsnm{Batzner}, \binits{S.}},
\bauthor{\bsnm{Musaelian}, \binits{A.}},
\bauthor{\bsnm{Sun}, \binits{L.}},
\bauthor{\bsnm{Geiger}, \binits{M.}},
\bauthor{\bsnm{Mailoa}, \binits{J.P.}},
\bauthor{\bsnm{Kornbluth}, \binits{M.}},
\bauthor{\bsnm{Molinari}, \binits{N.}},
\bauthor{\bsnm{Smidt}, \binits{T.E.}},
\bauthor{\bsnm{Kozinsky}, \binits{B.}}:
\batitle{E(3)-equivariant graph neural networks for data-efficient and accurate interatomic potentials}.
\bjtitle{Nature Communications}
\bvolume{13}(\bissue{1}),
\bfpage{2453}
(\byear{2022})
\doiurl{10.1038/s41467-022-29939-5}
\end{barticle}
\endbibitem

\bibitem[\protect\citeauthoryear{Li et~al.}{2020}]{li_unified_2020}
\begin{barticle}
\bauthor{\bsnm{Li}, \binits{R.}},
\bauthor{\bsnm{Lee}, \binits{E.}},
\bauthor{\bsnm{Luo}, \binits{T.}}:
\batitle{A unified deep neural network potential capable of predicting thermal conductivity of silicon in different phases}.
\bjtitle{Materials Today Physics}
\bvolume{12},
\bfpage{100181}
(\byear{2020})
\doiurl{10.1016/j.mtphys.2020.100181}
\end{barticle}
\endbibitem

\bibitem[\protect\citeauthoryear{Bhullar et~al.}{2024}]{https://doi.org/10.1002/cphc.202400090}
\begin{barticle}
\bauthor{\bsnm{Bhullar}, \binits{M.}},
\bauthor{\bsnm{Bai}, \binits{Z.}},
\bauthor{\bsnm{Akinpelu}, \binits{A.}},
\bauthor{\bsnm{Yao}, \binits{Y.}}:
\batitle{Phase transition in silicon from machine learning informed metadynamics}.
\bjtitle{ChemPhysChem}
\bvolume{25}(\bissue{13}),
\bfpage{202400090}
(\byear{2024})
\doiurl{10.1002/cphc.202400090}
{\href{https://arxiv.org/abs/https://chemistry-europe.onlinelibrary.wiley.com/doi/pdf/10.1002/cphc.202400090}{{https://chemistry-europe.onlinelibrary.wiley.com/doi/pdf/10.1002/cphc.202400090}}}
\end{barticle}
\endbibitem

\bibitem[\protect\citeauthoryear{Mulliken}{1955}]{mullikenElectronicPopulationAnalysis1955}
\begin{barticle}
\bauthor{\bsnm{Mulliken}, \binits{R.S.}}:
\batitle{Electronic {{Population Analysis}} on {{LCAO}}--{{MO Molecular Wave Functions}}. {{I}}}.
\bjtitle{The Journal of Chemical Physics}
\bvolume{23}(\bissue{10}),
\bfpage{1833}--\blpage{1840}
(\byear{1955})
\doiurl{10.1063/1.1740588}
\end{barticle}
\endbibitem

\bibitem[\protect\citeauthoryear{Bader}{1990}]{10.1093/oso/9780198551683.001.0001}
\begin{bbook}
\bauthor{\bsnm{Bader}, \binits{R.F.W.}}:
\bbtitle{Atoms in Molecules: A Quantum Theory}.
\bpublisher{Oxford University Press},
\blocation{Oxford, UK}
(\byear{1990}).
\doiurl{10.1093/oso/9780198551683.001.0001}
\end{bbook}
\endbibitem

\bibitem[\protect\citeauthoryear{Kresse and Furthm{\"u}ller}{1996}]{PhysRevB.54.11169}
\begin{barticle}
\bauthor{\bsnm{Kresse}, \binits{G.}},
\bauthor{\bsnm{Furthm{\"u}ller}, \binits{J.}}:
\batitle{Efficient iterative schemes for ab initio total-energy calculations using a plane-wave basis set}.
\bjtitle{Physical Review B}
\bvolume{54}(\bissue{16}),
\bfpage{11169}--\blpage{11186}
(\byear{1996})
\doiurl{10.1103/PhysRevB.54.11169}
\end{barticle}
\endbibitem

\bibitem[\protect\citeauthoryear{Perdew et~al.}{1996}]{PhysRevLett.77.3865}
\begin{barticle}
\bauthor{\bsnm{Perdew}, \binits{J.P.}},
\bauthor{\bsnm{Burke}, \binits{K.}},
\bauthor{\bsnm{Ernzerhof}, \binits{M.}}:
\batitle{Generalized gradient approximation made simple}.
\bjtitle{Physical Review Letters}
\bvolume{77}(\bissue{18}),
\bfpage{3865}--\blpage{3868}
(\byear{1996})
\doiurl{10.1103/PhysRevLett.77.3865}
\end{barticle}
\endbibitem

\bibitem[\protect\citeauthoryear{Kresse and Joubert}{1999}]{PhysRevB.59.1758}
\begin{barticle}
\bauthor{\bsnm{Kresse}, \binits{G.}},
\bauthor{\bsnm{Joubert}, \binits{D.}}:
\batitle{From ultrasoft pseudopotentials to the projector augmented-wave method}.
\bjtitle{Physical Review B}
\bvolume{59}(\bissue{3}),
\bfpage{1758}--\blpage{1775}
(\byear{1999})
\doiurl{10.1103/PhysRevB.59.1758}
\end{barticle}
\endbibitem

\bibitem[\protect\citeauthoryear{Wang et~al.}{2021}]{VASPKIT}
\begin{barticle}
\bauthor{\bsnm{Wang}, \binits{V.}},
\bauthor{\bsnm{Xu}, \binits{N.}},
\bauthor{\bsnm{Liu}, \binits{J.-C.}},
\bauthor{\bsnm{Tang}, \binits{G.}},
\bauthor{\bsnm{Geng}, \binits{W.-T.}}:
\batitle{Vaspkit: A user-friendly interface facilitating high-throughput computing and analysis using vasp code}.
\bjtitle{Computer Physics Communications}
\bvolume{267},
\bfpage{108033}
(\byear{2021})
\doiurl{10.1016/j.cpc.2021.108033}
\end{barticle}
\endbibitem

\bibitem[\protect\citeauthoryear{Thomas et~al.}{2018}]{thomas2018tensorfieldnetworksrotation}
\begin{botherref}
\oauthor{\bsnm{Thomas}, \binits{N.}},
\oauthor{\bsnm{Smidt}, \binits{T.}},
\oauthor{\bsnm{Kearnes}, \binits{S.}},
\oauthor{\bsnm{Yang}, \binits{L.}},
\oauthor{\bsnm{Li}, \binits{L.}},
\oauthor{\bsnm{Kohlhoff}, \binits{K.}},
\oauthor{\bsnm{Riley}, \binits{P.}}:
Tensor field networks: Rotation- and translation-equivariant neural networks for 3D point clouds
(2018).
\url{https://arxiv.org/abs/1802.08219}
\end{botherref}
\endbibitem

\end{thebibliography}

\end{document}